\documentclass[twocolumn,showpacs,preprintnumbers,amsmath,amssymb]{revtex4}

\usepackage{graphicx}% Include figure files
\usepackage{dcolumn}% Align table columns on decimal point

\usepackage{mathptmx, courier, pifont}
\usepackage[scaled=0.92]{helvet}
\usepackage[T1]{fontenc}
\usepackage{textcomp}

\usepackage{bm}% bold math

\newcommand{\singlefig}[6]{%
\begin{figure}\vspace{#3}%
\includegraphics*[scale=#5]{#2}%
\caption{\label{fig:#1} #6}%
\vspace{#4}%
\end{figure}}

\newcommand{\tsub}[1]{_{\mbox{\scriptsize#1}}}

\newcommand{\atanh}{{\rm atanh}}

\newcommand{\singletGap}{\Delta\tsub s}
\newcommand{\tripletGap}{\Delta\tsub t}

\newcommand{\mwgalign}[1]
{\hspace*{-0.4em}&#1&\hspace*{-0.4em}}  %%-->  Use in array and eqnarray
\newcommand{\fig}[1]{%
\ifx\labelshow\yes%                               % If \showlabels
Fig.~{ \{#1\}} \ref{fig:#1}%
\else%
Fig.~\ref{fg:#1}%                    % If \noshowlabels(default)
\fi}

\begin{document}

\title{A Unified Description of Cuprate and Iron Arsenide Superconductors}

\author{Mike Guidry$^{(1)}$}
%\email{guidry@utk.edu}
\author{Yang Sun$^{(2)}$}
%\email{ysun@nd.edu}
\author{Cheng-Li Wu$^{(3)}$}
%\email{clwu@phys.cts.nthu.edu.tw}

\affiliation{
$^{(1)}$Department of Physics and Astronomy, University of
Tennessee, Knoxville, Tennessee 37996, USA \\
$^{(2)}$Department of Physics, Shanghai Jiao Tong
 University, Shanghai 200240, People's Republic of China \\
$^{(3)}$Department of Physics, Chung-Yuan Christian University,
Chungli, Taiwan 320, ROC
}

\date{\today}

\begin{abstract}
We propose a unified description of cuprate and iron-based
superconductivity.  Consistency with  magnetic structure inferred
from neutron scattering implies significant constraints on the
symmetry of the pairing gap for the iron-based superconductors. We
find that this unification {\em requires} the orbital pairing
formfactors for the iron arsenides to differ fundamentally from
those for cuprates at the microscopic level.
\end{abstract}


\maketitle

\subsection{Introduction}

In a 2004 paper \cite{guid04} we proposed that an SU(4) dynamical
symmetry introduced in Ref.~\cite{guid99} had two properties
important for understanding high-temperature superconductivity (SC).
The first was that the SU(4) algebra imposed no double occupancy  by
symmetry, not projection. Thus  superconductivity emerges naturally
from an antiferromagnetic (AF) Mott insulator state at half filling.
The second was that SU(4) symmetry alone is sufficient to guarantee
many essential features of cuprate superconductivity, irrespective
of microscopic details such as pairing formfactors (except to the
extent that these  are broadly consistent with an emergent SU(4)
symmetry).

This led us to propose that cuprate superconductivity was a new kind
of superconductivity characterized by more complex behavior than
normal BCS superconductivity because the symmetry structure
associated with the superconductivity was non-abelian. The physical
content of this mathematical statement is that the non-abelian
algebra imposes dynamical constraints on the interaction of
collective degrees of freedom such as magnetism and charge with
superconductivity.  Because of the key dynamical role played by the
commutators, we termed this behavior {\em non-abelian
superconductivity}.

We demonstrated in Ref.\ \cite{guid04}, and amplified in  more
recent papers \cite{sun05,sun06,sun07,guid07,guid08b,guid08c}, that
{\em any} microscopic structure consistent with an algebra having
non-abelian subalgebras can lead to the complex behavior observed
for cuprate superconductors. In Ref.~\cite{guid04} we predicted that
there could be other compounds rather different from cuprate
superconductors in microscopic details that could exhibit properties
analogous to cuprate superconductors, provided that they realized in
their emergent properties a symmetry, such as SU(4), having
non-abelian subgroups and thus non-trivial commutators between
pairing and other degrees of freedom.

In early 2008 a series of experiments initiated in Japan and China
demonstrated a surprising new class of high-temperature
superconductors based on iron arsenides
\cite{kam06,wat07,kam08,che08,wen08,che08b,che08c,ren08,ren08b,che08d}.
These compounds have achieved critical temperatures $T\tsub c \sim
55$ K that are surpassed only by cuprates. These Fe-based
superconductors have an atomic structure differing from that of the
cuprates in significant details, yet there are many similarities
when compared with the cuprates.  This has led to a flurry of effort
to determine whether these two classes of superconductors share a
similar origin. At stake is the mechanism for Fe-based
superconductivity, but perhaps  a deeper understanding of that for
cuprate superconductivity as well.

In understanding the cuprates a key role was played by the
realization that superconductivity is dominated by singlet $d$-wave
pairs.  Hence, a major emphasis for the new Fe-based superconductors
has been to determine the symmetry of the pairing gap.
Experimentally the situation remains somewhat unclear. There is
substantial evidence that the pairing gap is spin singlet
\cite{gra08,mat08,kaw08}, but the gap orbital symmetry remains
unsettled.  Many experiments suggest that there are no gap nodes on
the Fermi surfaces but some find evidence for nodes
\cite{nak08,shan08,gang08,renc08,ahil08,naka08,gra08,wang09,muku08,mill08,wangXL08,hash09,kond08,ding08,chen08,park08,mu09},
suggesting that there may be more than one orbital gap symmetry
playing a role in the FeAs compounds.

Many theoretical proposals have been made  for the gap symmetry
(see, for example, Refs.\
\cite{aok08,xu08,kuro08,dai08,lee08,si08,yao08,qi08,wang08,wan08,
seo08,shi08,zhou08,you08,par08}). These typically start from
assumptions about microscopic structure and interactions near the
Fermi surface and attempt to predict the likely orbital and spin
structure for pairs.  This is a complex problem and different
authors reach different conclusions concerning the gap symmetry.

In this paper we propose that the  Fe-based superconductors are a
second example (after cuprates) of the non-abelian superconductivity
predicted in Ref.~\cite{guid04}.  We further propose that the
similarity of cuprate and Fe-based phenomenology indicates that
non-abelian superconductivity for the  FeAs superconductors is based
on the {\em same} SU(4) symmetry that explains the phenomenology of
the cuprates, thus providing a unified picture of Cu and Fe based
superconductors.  We argue that this is the case even if the
microscopic pairing mechanism in the two cases were to turn out to
be different.  Indeed, we shall provide evidence that unification at
the emergent degrees of freedom level is possible only if the two
classes of high temperature superconductors have {\em different}
microscopic pairing structure. Finally, we use the non-abelian
symmetry to {\em predict} the orbital and spin symmetries possible
for the Fe-based compounds by requiring consistency between observed
pairing and magnetic properties.

\subsection{Cuprate and Fe-Based Phenomenology}

Many experiments suggest strong similarities between cuprate and
Fe-based superconductors.  Both appear to be mediated by
electron--electron correlations rather than phonons (some dispute
this), and to involve the close proximity of antiferromagnetism (AF)
and superconductivity (SC).  In both the superconductivity  occurs
often (but not always) in 2-dimensional conducting planes and
corresponds to superconductors with low carrier density, and the
superconductors emerge from the parent compounds upon either hole or
particle doping from donor planes. On the other hand, some things
seem rather different between the two classes of superconductors.
Specifically, if we wish to address how similar the cuprate and FeAs
superconductors are the following issues are perhaps relevant:

\paragraph{Multiband Physics}

There is uniform agreement that FeAs superconductivity is multiband
and many think that this is crucial to the physics. The main
disagreement is over whether one must treat all five Fe bands near
the Fermi surface, or whether a simplified model with say two bands
captures most of the physics. Thus, there are multiple sheets for
the Fermi surface and the microscopic pairing formfactor for an
$n$-band model is actually an $n\times n$ matrix.

\paragraph{Multigap Physics}

Related to the question of multiband physics is the question of
multigap physics.  From the microscopic point of view, if there are
multiple bands near the Fermi surface that can contribute to the
pairing interaction the orbital formfactor for the pairing gap
becomes a matrix, which implies that there can be more than one
observable pairing gap. Various experiments in the FeAs compounds
see evidence for two or more pairing gaps, differing in size by as
much as a factor of two. For example, the ARPES measurements of
Ref.~\cite{nak08} find evidence for four Fermi surface sheets in
Ba$_{0.6}$K$_{0.4}$Fe$_2$As$_2$, and four corresponding gaps, with
the largest and smallest gaps differing by about a factor of two in
size.

\paragraph{Singlet Gap or Triplet Gap}

On general grounds we would expect that the pairing gap could be
either spin singlet or triplet, depending on the nature of the
appropriate effective interaction.  This is supported by the
numerical calculations of the Moreo--Dagotto group
\cite{dag08,mor09}, which find that either singlet or triplet pairs
could be favored energetically depending on the detailed
interactions.  However, NMR data now indicate that the FeAs pairing
gap has singlet spin character \cite{gra08,mat08,kaw08}.  Therefore,
we shall assume the observed gaps to be dominantly singlet, as for
cuprates.

\paragraph{Two-Dimensional Physics}

Another potential difference between FeAs compounds and cuprates
concerns whether the physics responsible for SC is dominated by
two-dimensional $a$--$b$ plane physics.  For the cuprates this
assumption is a relatively good approximation.  In the iron arsenide
compounds the situation is less clear.  The present evidence
suggests that for some FeAs superconductors the SC is rather
two-dimensional but for others the superconductivity depends
significantly on properties in the $c$-axis direction.
% the ``111'' compounds (xFeAs) are highly two-dimensional but the ``122'' compounds (xFe$_2$As$_2$) depend significantly on properties in the $c$-axis direction.

\paragraph{On-Site Coulomb Repulsion}

In the cuprates the large on-site Coulomb repulsion strongly
suppresses double occupancy and leads to a Mott insulator normal
state. In Ref.~\cite{guid04} we demonstrated that non double
occupancy for pairs in the real space is a sufficient condition to
guarantee that the minimal closed algebra is SU(4), independent of
detailed microscopic considerations such as the orbital or spin
symmetry of the gap and the associated structure of the pairs.

The situation in the FeAs compounds is less clear.  A variety of
calculations and arguments suggest that $U$ must lie in an
intermediate range between no correlations and the strong on-site
repulsion observed for the cuprates: if $U$ were too large the
parent states would develop a charge gap and be good insulators; if
it were too small the parent states would be good metals.  That they
are in fact found to be poor metals suggests an intermediate range
of $U$.  The presence experimentally of the spin density wave state
discussed below also implies a correlated metal, suggesting a
non-trivial $U$.

Various calculations indicate that the on-site repulsion on the Fe
atoms is approximately half of that observed for Cu atoms in the
cuprates, and the normal states for the iron arsenides are observed
to be AF metals, not AF Mott insulators.  However, these metals are
poor metals and various considerations suggest that the FeAs normal
states may generally be near a Mott transition. For example, an
analysis based on density functional and dynamical mean field theory
\cite{hau08} concludes that a realistic on-site Coulomb repulsion $U
\simeq 4$ eV would be sufficient to open a Mott gap for a single
band at the Fermi surface, but not for the five Fe bands expected to
be near the Fermi surface in FeAs undoped compounds.  Instead these
calculations give correlated metal structure but with poor charge
transport properties (a scattering rate at the Fermi level
corresponding to 0.4 eV at $T=116$ K).  However, a small increase of
the on-site repulsion to $U \simeq 4.5$ eV in these same
calculations begins to open a semiconductor-like gap even at room
temperature and Ref.~\cite{hau08} suggests that these compounds are
near the metal--insulator transition.

\paragraph{Nearest Neighbor or Next Nearest Neighbor Pairing}

Because the arsenic atoms are out of the Fe plane, general arguments
imply that next nearest neighbor (NNN) interactions can compete or
even exceed nearest neighbor (NN) interactions, raising the question
of whether possible bond-wise pairs (pairs with particles on
different lattice sites) involve NN or NNN.  Calculations indicate
that  for low enough values of the Hubbard repulsion $U$ the NN
pairing dominates the NNN pairing, but for increasing values of $U$
it is found that NNN pairing begins to compete more favorably with
NN  \cite{mor09}.

\paragraph{Differences in Antiferromagnetism}

Neutron scattering measurements indicate that for both cuprates and
pnictides antiferromagnetism is important and in close proximity to
the superconductivity in the phase diagram.  However, the nature of
the antiferromagnetism is different in the two cases. The schematic
spin structure associated with the undoped FeAs compounds that is
consistent with neutron scattering results is illustrated in the bottom portion of Fig.~\ref{fig:spinPlane}.%
 \singlefig
     {spinPlane}
     {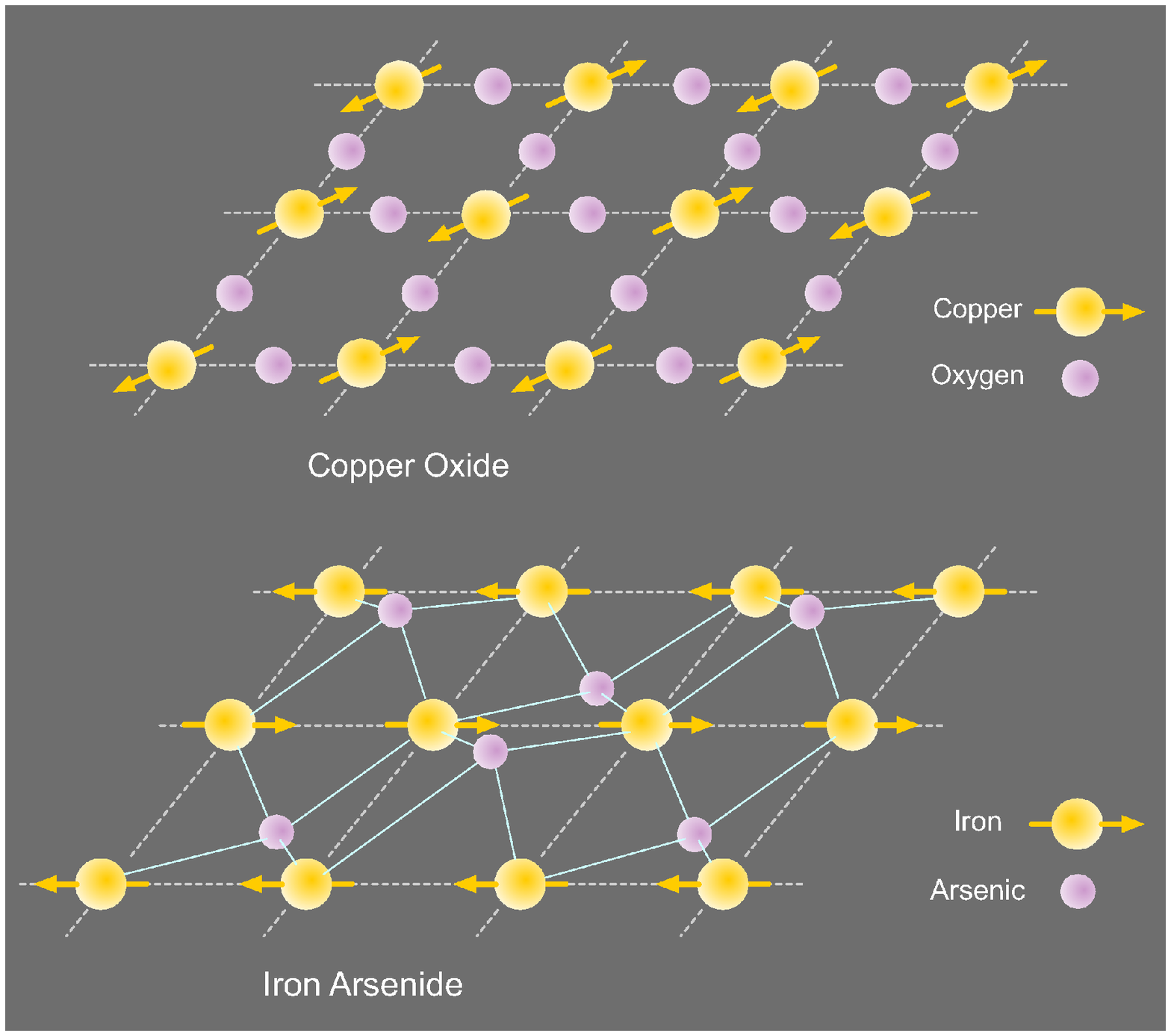}
     {0pt}
     {0pt}
     {0.46}
     {Schematic spin structure for cuprate and FeAs compounds.
The undoped iron arsenides are characterized by a
``stripe antiferromagnetism'' magnetic structure that differs from the
AF observed in the cuprates.}
This structure corresponds to alternating spins in one  direction
but stripes with spins all aligned with each other in the orthogonal
direction \cite{don08,cru08}.  This is in contrast to the
antiferromagnetism of the cuprate parent compounds illustrated in
the upper portion of Fig.~\ref{fig:spinPlane}, where in either the
horizontal or vertical direction the spins alternate.  It is with
respect to such a magnetic structure that we must add or remove
electrons to make FeAs superconductors, if we adopt the point of
view that the magnetism and superconductivity are closely related
and that the latter develops out of the former with doping.

\subsection{Similarities and Differences}

We shall now make a case that the {\em similarities} observed for
cuprate and Fe-based SC suggest that a minimal description of either
involves the same SU(4) Lie algebra, and that this algebra is
sufficient to ensure a unified picture of the most general
properties observed for these superconductors. Further, we shall
argue that the {\em differences} between the two classes of
superconductors do not change the algebraic structure of a minimal
model and thus do not change the most fundamental properties of
these superconductors, but rather influence the theory only
parametrically. Thus, we shall propose a unification in the emergent
degrees of freedom that can exist for cuprate and FeAs
superconductors, even if the microscopic structure and the pairing
gap symmetry are different in the two cases.

\subsection{A Minimal Closed Set of Operators}

Let us use the phenomenology of the Fe-based superconductors to
infer a minimal set of operators consistent with the observed
degrees of freedom.  It is clear that antiferromagnetism lies in
close proximity to superconductivity in the phase diagram.  Thus, we
require AF and pairing operators. Unlike for cuprates where the AF
operator has a lattice vector $\bm Q = (\pi,\pi)$, in the iron
arsenides neutron scattering experiments indicate that the AF
lattice vector is $\bm Q = (0, \pi)$ or $(\pi,0)$
\cite{don08,cru08}, reflecting the difference in spin structure
illustrated in Fig.~\ref{fig:spinPlane}.  Considerable attention has
been paid to whether the pairing gap symmetry is spin singlet or
triplet, with recent Knight shift data favoring singlet pairing
\cite{gra08,mat08,kaw08}.  However, by a similar argument as for the
cuprate superconductors \cite{guid08b}, in the presence of an AF
field the theory must admit both singlet and triplet pairs in a
self-consistent Hilbert space, since the AF interactions can scatter
singlet pairs into triplet pairs and vice versa. Physically, one or
the other kind of pair may dominate energetically and in the
superconducting charge transport, but quantum mechanically both
kinds of pairs must be permitted if one kind is.  Finally the system
has a conserved charge and spin, so we require operators for these
also.

The presence of strongly competing AF and SC ensures that Fe-based
superconductors correspond to a non-abelian symmetry as discussed in
Ref.~\cite{guid04}, but the relevant non-abelian algebra need not be
the SU(4) cuprate algebra. The question of the minimal closed
algebra for the iron arsenides turns on whether the pairing that can
produce the superconductivity involves both on-site pairs (two
particles on the same lattice site) and bondwise pairs (two
particles on different lattice sites).  If we make the simplest
model and restrict attention to a single kind of nearest neighbor or
next nearest neighbor bondwise pair, the arguments of
Ref.~\cite{guid04} indicate that the minimal closed algebra is SU(4)
if there are only bondwise (no on-site) pairs.

For the cuprates the on-site Coulomb repulsion is very strong, which
opens a substantial gap between the bondwise and on-site pairs.
Thus it is a very good approximation to assume that only bondwise
pairs contribute to the ground state properties at low temperature.
For the iron arsenides the situation is less clear but, as we have
noted, many investigations suggest on-site repulsion that is of
intermediate strength relative to the cuprates.  However, in the
present context the issue is not whether the on-site repulsion is
strong enough to suppress double occupancy in general, but only
whether the correlations are sufficient to push on-site collective
pairs to substantially higher energy than bondwise collective pairs.
If that is the case, then we may construct a minimal low-energy
model involving only bondwise pairs.  If these pairs do not overlap
on the spatial lattice (that is, the wavefunction is a superposition
of bondwise pairs where no lattice site is occupied by particles
from two different collective pairs), then the arguments of
Ref.~\cite{guid04} imply that the resulting closed algebra is SU(4).

Since this is a rather significant point, it is worth further
elaboration.  In Ref.~\cite{guid04} we emphasized the important
relationship between no double occupancy of the lattice and SU(4)
symmetry.  While that discussion was adequate for the basics of
cuprate superconductivity where the parent compounds are strong
insulators, it is necessary to express it more precisely when
dealing with more general cases of non-abelian superconductivity
where the parent compounds might be insulators, metals, or poor
metals.

The key distinction to make is between the coherent pairs of the
SU(4) symmetry-truncated basis (which are responsible for charge
transport in the superconducting state) and additional valence
particles that are not part of the coherent pairs and are
responsible for charge transport in the normal state at zero
temperature. Thus the statement in Ref.~\cite{guid04} that closure
of the SU(4) algebra requires the lattice to not be doubly occupied
is a statement specifically about site occupation by {\em components
of coherent pairs:} closure of the SU(4) algebra requires (1) that
there be no collective on-site pairs and (2) that no collective
bondwise pairs overlap on the spatial lattice.

Thus closure of the SU(4) Lie algebra respects the Mott insulator
characteristics of the cuprate normal states in that it represents
optimal configurations for competing AF and SC in the presence of
strong on-site repulsion.  However, the SU(4) solution in general
{\em need not correspond to an insulator,} since the (normal-state)
charge transport properties are determined by the properties of
particles not in the collective pairs.  For example, the on-site
repulsion could be sufficiently strong to make it energetically
unfavorable either  to form on-site collective pairs or to have
double site occupancy by unpaired particles, in which case the
symmetry is SU(4) and in addition the material would be expected to
be insulating.  This is representative of the situation in the
cuprates.  But one could also have a situation where the same-site
repulsion strongly disfavors on-site pairs over bondwise pairs, but
does not forbid some charge transport by the unpaired particles in
the normal state.  The resulting material would again be described
by SU(4), but now is expected to be a metal (or poor metal).  This
situation is representative of the FeAs compounds.

Although a few models involving on-site pairs have been proposed for
the iron arsenides, numerical calculations such as those of
Refs.~\cite{dag08,mor09} indicate that the dominant pairing channels
involve nearest neighbor or next nearest neighbor {\em bondwise
pairing}.  We take this as microscopic evidence that a minimal
description of FeAs high temperature superconductors involves
dominantly bondwise pairing for physically reasonable ranges of the
on-site repulsion $U$ and thus corresponds to an SU(4) symmetry.

We conclude from the preceding discussion that the pnictides exhibit
a form of non-abelian superconductivity that is in fact  {\em
isomorphic (described by the same SU(4) algebra) to the nonabelian
superconductivity of the cuprates.} This isomorphism and
corresponding similarity of the Fe-based and cuprate phenomenologies
suggests two unique opportunities. (1)~If the algebra associated
with the minimal set of Fe-based operators closes under commutation
then both cuprate and Fe-based superconductivity can be described by
SU(4) symmetry, thus providing a unified description of these two
types of superconductivity, irrespective of differences at a
microscopic level. (2)~Requiring the algebra to close implies a set
of non-linear equations on the operators that, coupled with data,
can be used to constrain their microscopic form.  In particular,
these relations can be used to {\em predict} the allowed orbital
symmetries of the pair gap by requiring consistency of those gaps
with the observed magnetic structure. In the following we shall
implement these ideas quantitatively.

\subsection{Extension of SU(4) to Multiband Pairing}

The preceding general arguments suggest that the nonabelian
superconductivity found in the iron arsenides corresponds to an
SU(4) symmetry.  To implement this idea quantitatively, we must
extend the SU(4) formalism developed for the cuprates under the
simplifying assumption of a single band near the Fermi surface to
the case where multiple bands may lie near the Fermi surface within
the Brillouin zone. With the preceding discussion and
Refs.~\cite{guid99} as a guide, we introduce the following set of
operators:
\begin{subequations}
 \label{E1}
\begin{eqnarray}
p^\dagger \mwgalign= \sum_{\bm k b b'} g(\bm k) \alpha_{\bm k b}
\alpha_{\bm -k b'}
c_{\bm k b \uparrow}^\dagger
c_{-\bm k b' \downarrow}^\dagger
\quad p=(p^\dagger)^\dagger
\label{E1.1}
\\
q_{ij}^\dagger \mwgalign= \sum_{\bm k b b'} g(\bm k) \alpha_{\bm k + \bm Q, b}
\alpha_{\bm -k b'}
c_{\bm k+\bm Q,b i}^\dagger c_{-\bm k,b' j}^\dagger
\quad q = (q^\dagger)^\dagger
\label{E1.2}
\\
Q_{ij} \mwgalign= \sum_{\bm k b b'} \alpha_{\bm k + \bm Q, b} \alpha^*_{\bm k b'}
c_{\bm k+\bm Q,b i}^\dagger c_{\bm k b' j}
\label{E1.3}
\\
 S_{ij} \mwgalign=
\sum_{\bm k b b'} \alpha_{\bm k b} \alpha_{\bm k b'}^*
c_{\bm k,b i}^\dagger c_{\bm k,b' j} - \frac12 \Omega \delta_{ij}
\label{E1.4}
\end{eqnarray}
\end{subequations}
where $\alpha_{\bm kb}$ is the amplitude to have a $b$-band electron with momentum
$\bm k$,
 $c_{\bm k,b, i}^\dagger$ creates a fermion of momentum $\bm k$ and
spin projection $i,j= 1 {\rm\ or\ }2 = \ \uparrow$ or $\downarrow$
in band $b$, $\bm Q$ is an AF ordering vector, $\Omega$ is the
effective lattice degeneracy, which is the maximum allowed number of
doped electrons that can form coherent SU(4) pairs, and $g(\bm k)$
is a pairing formfactor.  These operators are equivalent to the
SU(4) generators defined in Refs.~\cite{guid99} except that the
fermion operators have acquired a band index $b$, with the number of
bands included in the sum determined by the physics of the
particular problem being addressed. If we define effective one-band
creation and annihilation operators through
\begin{equation}
 a^\dagger_{ki} = \sum _b \alpha_{\bm k b} c^\dagger_{\bm k b i}
\qquad
a_{\bm k i} = (a^\dagger_{ki})^\dagger
\qquad
\sum_b |\alpha_{\bm k b}|^2 = 1,
\label{oneBandEff}
\end{equation}
then Eqs.~(\ref{E1}) may we written as
\begin{subequations}
\label{E3}
\begin{eqnarray}
p^\dagger\mwgalign=\sum_{\bm k} g(\bm k) a_{\bm k\uparrow}^\dagger
a_{-\bm k\downarrow}^\dagger
\qquad p=(p^\dagger)^\dagger
\label{E3.1}
\\
q_{ij}^\dagger \mwgalign= \sum_{\bm k} g(\bm k) a_{\bm k+\bm Q,i}^\dagger a_{-\bm
k,j}^\dagger
\qquad q = (q^\dagger)^\dagger
\label{E3.2}
\\
Q_{ij} \mwgalign= \sum_{\bm k} a_{\bm k+\bm Q,i}^\dagger a_{\bm k,j} \qquad S_{ij} =
\sum_{\bm k}
a_{\bm k,i}^\dagger a_{\bm k,j} - \frac12 \Omega \delta_{ij} ,
\label{E3.3}
\end{eqnarray}
\end{subequations}
which is now exactly the form of the  U(4) $\supset$ U(1) $\times$
SU(4) generators discussed originally in Refs.~\cite{guid99}.

Inserting the AF ordering vector $\bm Q = (Q_x, Q_y) = (0, \pi)$
appropriate for the observed FeAs magnetic structure and calculating
all commutators for the operators (\ref{E1}), we find that the set
is closed under commutation (returns a linear combination only of
this set of operators) provided that three conditions are satisfied
by the pair formfactor:
\begin{equation}
g(\bm k) = g(-\bm k)\qquad g(\bm k + \bm Q) = \pm g(\bm k)
\qquad |g(\bm k)| = 1.
\label{conditions}
\end{equation}
If these conditions are met, the operators defined in Eq.~(\ref{E1})
close a Lie algebra isomorphic to the U(4) $\supset$ U(1) $\times$
SU(4) algebra describing cuprate superconductivity.  By defining new
operators that are linear combinations of the operators (\ref{E1}),
we may restrict attention to the SU(4) subalgebra that is generated
by 15 operators corresponding physically to singlet pairing, triplet
pairing, antiferromagnetism, spin and charge \cite{guid99,guid04}.

Let's illustrate these ideas more graphically for two bands, with
the argument easily generalized to more bands. Simple possibilities
for forming SU(4) pairs  are illustrated schematically in Fig.\
\ref{fig:pairs2OrbitNN} for nearest neighbor pairs and
Fig.\ \ref{fig:pairs2OrbitNNN} for next nearest neighbor pairs,%
 \singlefig
     {pairs2OrbitNN}
     {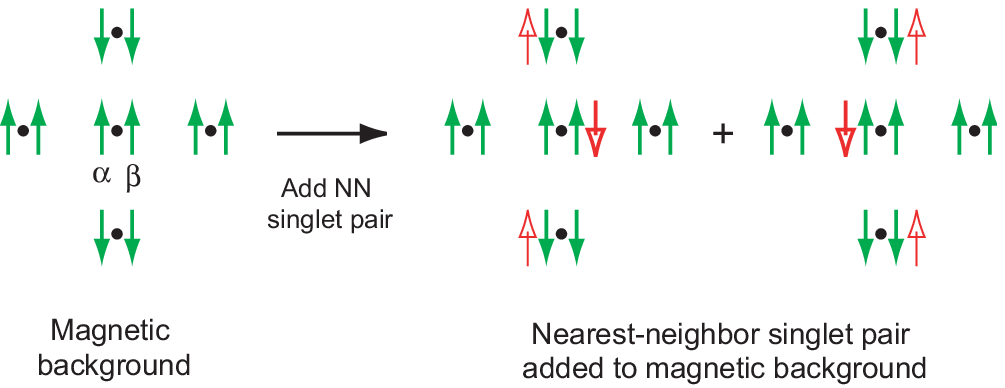}
     {0pt}
     {0pt}
     {0.84}
     {Possible SU(4) nearest neighbor (NN) spin-singlet
pair structure in a 2-orbital model for a pair centered on a
particular site. Arrows indicate spin-up and spin-down particles,
with an arrow to the left of a lattice point signifying a particle
in orbital $\alpha$ and an arrow to the right of a lattice point
signifying a particle in  orbital $\beta$. Solid green arrows
represent the undoped magnetic background state. The open red arrows
represent the added pair. The collective SU(4) NN pair would then
correspond to a coherent sum over the lattice of such pairs centered
on individual lattice sites.}
 \singlefig
     {pairs2OrbitNNN}
     {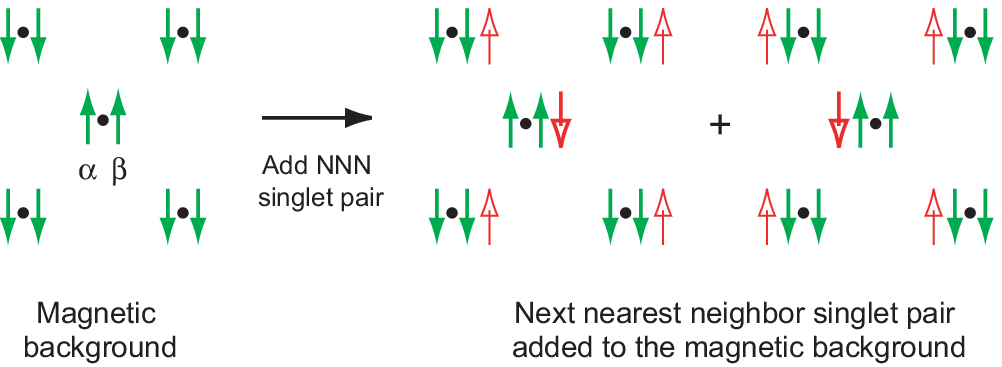}
     {0pt}
     {0pt}
     {0.84}
     {As for Fig.~\ref{fig:pairs2OrbitNN}
but for a possible SU(4) next nearest neighbor (NNN) spin-singlet
pair structure in a 2-orbital model for a pair centered on a
particular site. }
where now we concentrate on spin-singlet pairs  and where we have
assumed that the vacuum to which we add pairs is the stripe-AF
magnetic state found in neutron scattering experiments. In this
picture we see that the multiband pairs are formed in a manner
similar to the case for SU(4) in the single-band cuprates, but now
each individual pair in the coherent sum over the lattice is more
complex in structure because the electrons in the bond-wise pair are
distributed over more than one orbit on each site.
Figures~\ref{fig:pairs2OrbitNN} and \ref{fig:pairs2OrbitNNN} are of
course just cartoons illustrating the coherent multiband pairs
defined precisely in Eq.~(\ref{E1}).

Thus, as long as the pairs entering the formalism as basis states
may be viewed as strongly collective, multiband pairing in the
presence of antiferromagnetism may be treated formally in the same
way as single-band pairing in the presence of antiferromagnetism. Of
course the pairs may have a richer internal structure in the
multiband case and experiments sensitive to single-particle degrees
of freedom like ARPES may probe that structure.  But at the level of
emergent collective degrees of freedom and their corresponding
physical consequences, the important point is that the collective
pairs---whatever their detailed internal structure---close an
algebra when commuted with other physically relevant operators in
the system.  If that is the case, then the overall physics of the
problem is strongly constrained by the corresponding (generally
nonabelian) algebra and one has a theoretical formalism constructed
specifically to deal with the common general features of a class of
superconductors.

The philosophy that underlies this approach rests on the assumption
that a set of emergent strongly-correlated electron phenomena
observed to occur with some systematic phenomenology over a range of
compounds can be best understood by first identifying the features
common to all occurrences and viewing the differences as
perturbations around the unifying features.  This approach can fail
if there are no common recurring collective modes across a class of
compounds, or if the recurring collective modes are strongly
perturbed from compound to compound.  However, there is abundant
empirical evidence in the cuprates that this approach is a
physically reasonable starting point, and  in the FeAs compounds we
believe that the evidence is also strong enough to justify a focus
on the unifying emergent properties across compounds as a more
reasonable starting point than one focusing on the differences at
the microscopic level.

\subsection{Pair Formfactors and Closed Algebras}

Let us now consider in more detail the closure conditions
(\ref{conditions}). The first requirement $g(\bm k) = g(-\bm k)$ is
almost always satisfied by physically reasonable formfactors. As
discussed  in Ref.~\cite{guid04}, the condition $|g(\bm k)| = 1$
necessary to close the algebra in momentum space may be interpreted
as an occupation constraint on the {\em full formfactor without this
condition} in the real space. Specifically, for cuprates the
$d$-wave formfactor $g(\bm k) = \cos k_x - \cos k_y$ must be
approximated by ${\rm sgn\ } (\cos k_x -\sin k_x)$ to close the
algebra in momentum space.  However, if the operators are Fourier
transformed to the real space retaining the  {\em full formfactor}
(that is, $\cos k_x - \cos k_y$) the SU(4) algebra closes but only
if the lattice is restricted to no double pair occupancy. This
indicates that  $|g(\bm k)| = 1$ is not an approximation but rather
is a physically necessary momentum-space corollary to no double
occupancy (by pairs) for the collective wavefunction in the real
space. Therefore,  we shall assume that closure of the SU(4) algebra
requires no double site occupancy by pairs and the condition
\begin{equation}
 g(\bm k + \bm Q) = \pm g(\bm k)
\label{conditions2}
\end{equation}
applied to the full formfactor (without the condition $|g(\bm
k)|=1$), which will imply a set of real-space occupancy constraints
imposed by the algebra that depends on the exact form of $g(\bm k)$.

\subsection{Unified SU(4) Model}

Since Fe-based SC is observed to occur with many features similar to
that for cuprate SC, the simplest assumption is that the operator
set (\ref{E1}) describes the minimal collective degrees of freedom
consistent with iron arsenide phenomenology.  But this, coupled with
our provisional assumption of suppressed double site occupancy by
pairs, means that the minimal algebra consistent with FeAs
phenomenology is SU(4), just as for the cuprates. Furthermore, with
this assumption Eq.~(\ref{conditions2}) provides immediate
constraints on the permissable form of $g(\bm k)$ for Fe-based
compounds. In Table \ref{FeGapTable} we apply these constraints to
some gap symmetries that have been proposed for Fe-based
superconductors, indicating whether the SU(4) algebra can be closed
for each assumption, and for reference we carry out the same
procedure for the cuprates (without regard to whether each symmetry
has been proposed seriously for cuprates).  Some of the formfactors
considered in Table~\ref{FeGapTable} are illustrated in
Fig.~\ref{fig:contour4filled}.

% ---------------  Beginning of Table  --------------- %

\begin{table}[t]
  \centering
  \caption{Some pairing gap orbital symmetries,
whether they satisfy Eq.~(\ref{conditions2})
and thus
  close the SU(4) algebra for Fe-based and cuprates, and
maximum doping fraction $P_f$ for allowed
FeAs symmetries.}
  \label{FeGapTable}
\vspace{1pt}
  \begin{normalsize}
    \begin{centering}
      \setlength{\tabcolsep}{5 pt}
      \begin{tabular}{cccc}
        \hline
            $g(\bm k)$ &
            Fe-based &
            Cuprate & $P_f$

        \\[1 pt]        \hline
            $s_{x^2+y^2}=\cos k_x+\cos k_y$ &
            No &
            Yes & --

        \\[1 pt]
            $d_{x^2-y^2} = \cos k_x-\cos k_y$ &
            No &
            Yes & --

        \\[1 pt]
            $s_{x^2y^2}=\cos k_x\cos k_y$ &
            Yes &
            Yes & $1/3$

        \\[1 pt]
            $d_{xy} = \sin k_x\sin k_y $ &
            Yes &
            Yes & $1/3$
    \\[1 pt]
            $s_{x^2+y^2} \pm d_{x^2-y^2} $ &
            Yes &
            Yes & $2/3$

    \\[1 pt]
            $s_{x^2+y^2} \pm id_{x^2-y^2} $ &
            No &
            Yes & --

        \\[1 pt]        \hline
      \end{tabular}
    \end{centering}
  \end{normalsize}
\end{table}

% ---------------  End of Table  --------------- %

\singlefig {contour4filled} {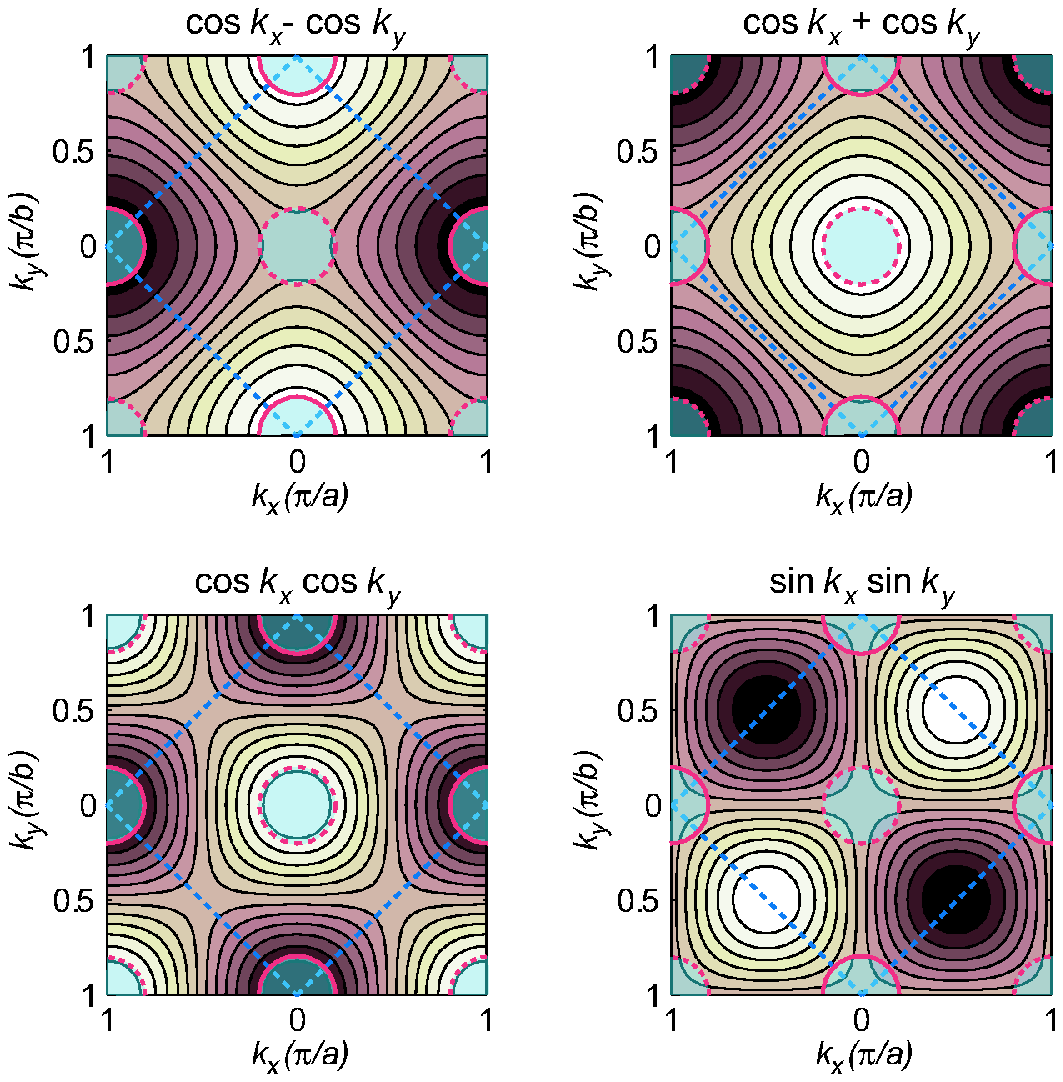} {0pt} {0pt}
{0.78} {Momentum-space formfactors for some  cases listed in
Table~\ref{FeGapTable}. The outer black square for each diagram is
the large Brillouin zone  associated with the Fe-only real-space
lattice. The dashed blue diamond inset in each box is the small
Brillouin zone associated with the true real-space lattice.
% Thus the outer black squares are the effective Brillouin zones and the inset blue
% diamonds are the true Brillouin zones.
Very schematic locations for the small Fermi surface pockets
obtained from typical calculations are sketched as heavy red curves:
solid for regions with electron pockets and dashed for regions with
hole pockets. }

From Table \ref{FeGapTable} we conclude that $s_{x^2+y^2}$ and
$d_{x^2-y^2}$ do not lead to a closed SU(4) algebra, but symmetries
such as $s_{x^2y^2}$ and $d_{xy}$ can. Thus we predict that neither
$s_{x^2+y^2}$ nor $d_{x^2-y^2}$ can be correct orbital symmetries
for the Fe-based superconductors because they are fundamentally
incompatible with the observed magnetic structure.  Hence a unified
SU(4) model of cuprate and Fe-based superconductors requires that
the corresponding orbital symmetry of the pair gap for Fe-based
compounds {\em cannot} be the symmetry $d_{x^2-y^2}$ found for the
cuprates. More generally, one can see from  Table \ref{FeGapTable}
that the more symmetric antiferromagnetism of the cuprates is
compatible with many possible pairing formfactors (though most
appear to not be realized physically), but the asymmetric AF of the
iron arsenides places much stronger constraints on a compatible
pairing structure.

It is easily verified that $\bm k \rightarrow \bm k + \bm Q$
interconverts the $s_{x^2+y^2}$ and $d_{x^2-y^2}$ formfactors, which
implies that while neither separately can close the SU(4) algebra
because they cannot satisfy Eq.~(\ref{conditions2}), the linear
combination
$$
g(\bm k) = g(s_{x^2+y^2}) \pm g(d_{x^2-y^2}),
$$
which is proportional to $\cos k_x$ or $\cos k_y$, does close the
algebra. On the other hand, the time-reversal breaking linear
combination
% \begin{equation}
$$
g(\bm k) = g(s_{x^2+y^2}) + i g(d_{x^2-y^2})
$$
% \label{linearCombo2}
% \end{equation}
proposed in Ref.\ \cite{lee08b} as a possible FeAs gap symmetry is
seen from Table \ref{FeGapTable} to be compatible with cuprate
antiferromagnetism but not with FeAs antiferromagnetism.

There is a further constraint that can be placed on the orbital
formfactor.  The coherent pair states corresponding to allowed
formfactors in Table \ref{FeGapTable} have a structure
\newcommand{\op}[1]{c^{\dagger}_{#1}}
$
D^\dagger = \sum_{r=\{x,y\}} \op{r\uparrow}
   \op{\bar r \downarrow}
$ (with $D^\dagger \equiv p^\dagger$) in the real space, where
$\op{r i}$ is the electron creation operator $a_{\bm k i}^\dagger$
in the coordinate representation. The $\op{\bar r i}$ for
$s_{x^2+y^2} \pm d_{x^2-y^2}$ pairs are formed from
nearest-neighbors:
\begin{subequations}
\begin{eqnarray}
 \op{\bar ri} \mwgalign= 2^{-1/2} (
\op{(x+a,y)i} + \op{(x-a,y)i} ) \qquad g(\bm k) = \cos k_x
\\
 \op{\bar ri} \mwgalign= 2^{-1/2} (
\op{(x,y+a)i} + \op{(x,y-a)i} ) \qquad g(\bm k) = \cos k_y ,
\end{eqnarray}
\end{subequations}
and for $\cos k_x \cos k_y$ or $\sin k_x \sin k_y$ pairs are formed
from next-nearest neighbors:
\begin{equation}
 \op{\bar ri} = {\scriptstyle\frac12} (
\op{(x+a,y+b)i} + \op{(x-a,y-b)i}
\pm \op{(x+a,y-b)i} \pm \op{(x-a,y+b)i} ),
\label{pairnnn}
\end{equation}
% \begin{eqnarray}
%  \op{\bar r\downarrow} \mwgalign= {\scriptstyle\frac12} (
% \op{(x+a,y+b)\downarrow} + \op{(x-a,y-b)\downarrow}
% \nonumber\\
% \flush\pm \op{(x+a,y-b)\downarrow} \pm \op{(x-a,y+b)\downarrow} ),
% \label{pairnnn}
% \end{eqnarray}
with $(+)$ corresponding to $\cos k_x \cos k_y$ and $(-)$ to $\sin
k_x \sin k_y$. These are illustrated in Fig.\
\ref{fig:primitivePair}.

\singlefig
{primitivePair}       % Label
{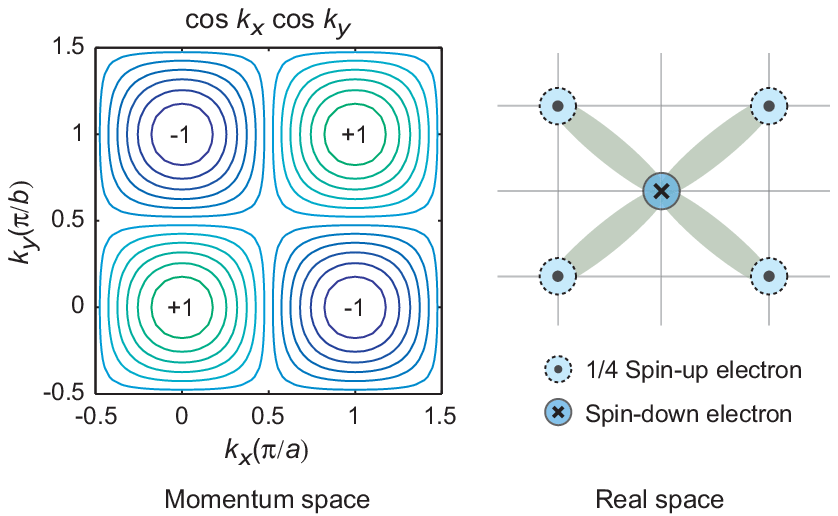}    % Filename (with path if needed)
{0pt}         % Extra space above (with units)
{0pt}         % Extra space below (with units)
{1.02}         % Scale factor (1.0 is full size)
{Pairing gap corresponding to a $\cos k_x \cos k_y$ formfactor in
momentum space and the corresponding schematic real-space pair
structure for a singlet electron pair. The shorthand notation ``$
1/4$ electron'' indicates  that the spin-up electron is distributed
with equal probability on four next-nearest neighbor sites in the
pair wavefunction of Eq.~(\ref{pairnnn}). The spatial pair structure
for $\sin k_x \sin k_y$ is similar to that for $\cos k_x \cos k_y$,
differing only in phases.}

As discussed in Ref.\ \cite{guid04} for cuprates, since SU(4)
symmetry requires no double occupancy by pairs there is a lattice
occupancy restriction associated with each of these pair structures.
By counting the maximum number of pairs that can be placed on the
lattice without overlap, as illustrated in Fig.~\ref{fig:pairCount},
the largest doping fraction consistent with SU(4) symmetry is found
to be $P_f = 2/3$ for $\cos k_x$ and $P_f = 1/3$ for $\cos k_x \cos
k_y$ or $\sin k_x \sin k_y$. These are summarized in the last column
of Table 1. Current data suggest that the superconductivity does not
extend much beyond $P_f = 1/3$. Even allowing for some uncertainty
in how many doped particles end up in the coherent pairs, this
strongly favors $\cos k_x \cos k_y$ or $\sin k_x \sin k_y$ among the
allowed orbital symmetries for iron arsenides in Table 1.

\singlefig{pairCount}{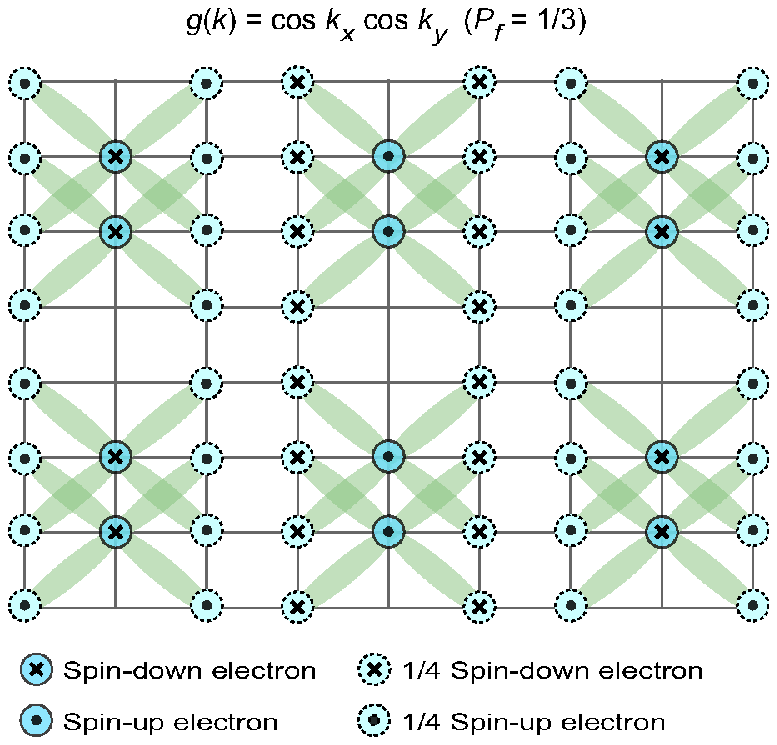}{0pt}{0pt}{1.0} {Schematic count
of maximum pair density consistent with SU(4) symmetry assuming
electron-doped material with a singlet $\cos k_x \cos k_y$ pair gap
formfactor. For this segment of the lattice, no additional pairs of
this structure can be added without causing a finite amplitude for
double site occupancy by pairs, which would break SU(4) symmetry.
By counting of occupied and unoccupied sites, the maximum fraction
of lattice sites that can be occupied by $\cos k_x \cos k_y$ pairs
without double occupancy is $1/3$.  The realistic wavefunction will
be a superposition of such configurations, each with a maximum pair
occupancy of $1/3$. The spatial pair structure and maximum doping
for $\sin k_x \sin k_y$ is the same as for $\cos k_x \cos k_y$,
since they differ only in phases (see Eq.~(\ref{pairnnn})).}

\subsection{Magnetism and Superconductivity}

The preceding discussion assumes implicitly that there is a
sufficiently intimate relationship between antiferromagnetism and
superconductivity in cuprate and FeAs superconductors that the
observed magnetic structure can be used to constrain the form of the
pairing interaction leading to the superconductivity.  There is
significant empirical evidence to suggest this. For example, recent
ARPES measurements in the underdoped pnictides by Xu et al
\cite{xu09} provide strong support for a picture very similar to the
one proposed theoretically in this paper.

However, it is by no means uniformly agreed (particularly in FeAs
compounds) that superconductivity and magnetism are so strongly
related and that the SC state develops directly from doping the AF
state. Thus we may invert the preceding discussion to provide a test
of this hypothesis: failure of the constraint predictions implied by
Table \ref{FeGapTable} would be strong evidence that AF and SC are
not sufficiently related in iron arsenides that one constrains the
form of the other. Conversely, verification of these predictions
would  support the SU(4) unification hypothesis for cuprate and FeAs
superconductivity proposed here, which would  imply that magnetism
and superconductivity are inextricably linked in high temperature
superconductors.

\subsection{Singlet and Triplet Pairing}

NMR measurements suggest singlet charge carriers for the FeAs
superconductors \cite{gra08,mat08,kaw08}. As discussed above,
consistency requires both singlet and triplet pairs in the truncated
SU(4) subspace and non-abelian superconductivity  can accommodate
either as charge carriers.  Which is realized depends on the
effective interaction, which can be determined empirically and could
differ between Fe-based and cuprate superconductors. For cuprates
the effective interaction is well determined and indicates that
singlet pair correlation is substantially larger than triplet pair
correlation \cite{sun05,sun06}. Because the singlet pairing energy
scale is much larger than the triplet pairing energy scale for
cuprate superconductors, we may expect that scattering at finite
temperature is larger for triplet pairs and that charge transport in
the superconducting state is dominated by coherent singlet pairs, as
observed. The understanding of iron arsensides is incomplete at this
point and the effective interaction is not as well established as
for cuprates, but the evidence that superconductivity in the iron
arsenides is also singlet in character suggests that similar
considerations apply for FeAs superconductivity.

\subsection{Spatial Inhomogeneity}

For the cuprates and (to a lesser degree) FeAs superconductors
there is evidence for a relatively universal phase diagram, but also
evidence for a variety of spatial inhomogeneity, particularly in the
underdoped region. As we have shown for the cuprates \cite{guid08c},
the coherent-state solutions for SU(4) have the natural property
that in underdoped compounds (and {\em only} in underdoped
compounds) there are many nearly degenerate ground states having
different ratios of pairing correlation to antiferromagnetic
correlation.  Thus the underdoped region is extremely sensitive to
external perturbation and a realistic wavefunction  may be expected
to be a superposition of components respecting SU(4) symmetry but
having different expectation values for pairing gaps and staggered
magnetization. (Mathematically, the wavefunction has components
corresponding to the same irreducible representations of SU(4),  but
to different irreducible representations with respect to its SO(4)
antiferromagnetic and SU(2) pairing subgroups.) This has two
important consequences:

\begin{enumerate}

 \item
It can produce a partially-gapped state above the superconducting
transition temperature in which there are significant correlations
corresponding to phase fluctuations modulated by competing AF and SC
order, but small expectation values for static order.

\item
This nearly-degenerate superposition of trial ground states will
exhibit a high degree of complexity (extreme susceptibility to
external perturbations) in the underdoped region.

\end{enumerate}

\noindent The first consequence leads to a quantitative description
of the pseudogap state (for which there is strong evidence in the
cuprates and growing evidence in FeAs compounds; see
Refs.~\cite{xu09,ish09,hai08,sat08,zha08}, for example); the second
implies that---{\em in the underdoped region only}---a rich variety
of spatial inhomogeneity can be induced by small background
perturbations, and that the nature of that inhomogeneity will depend
sensitively on the electronic and structural properties of
individual compounds.

Thus, the natural tendency to complexity exhibited by the SU(4)
solutions in the underdoped region, and the dependence of that
complexity on the detailed structure of individual compounds,
provide a possible explanation for differences in pseudogap behavior
and spatial inhomogeneity among different cuprates and pnictides.
Since the solution exhibits this complexity yet remains
SU(4)-symmetric in this emergent state, it can account for rich
variety in inhomogeneity within the (seemingly contradictory)
context of a global phase diagram.

\subsection{2D Versus 3D Physics}

We have implicitly or explicitly assumed 2D structure and the
corresponding language in much of our discussion because for
cuprates the 2D structure has strong experimental support.  But that
is not essential to our general approach:  if the theory is
formulated in momentum space the number of collective operators (and
thus the symmetry) need not change if the vector momentum indices on
the operators change their spatial dimensionality.  In that case the
physical interpretation of the underlying microscopy changes (with
attendant modification of the effective interaction operating in the
highly-truncated space), but the emergent symmetry governing the
global collective properties does not.

It has often been claimed that the effective low dimensionality of
the cuprate superconductivity (actual 3D in the AF N\'eel state
$\rightarrow$ effective 2D in the SC state) is a crucial ingredient
in the high-$T\tsub c$ mechanism.  For example, the underlying
philosophy of RVB and RVB-inspired theories is topological
charge--spin separation (spinons and holons), but the theories known
to exhibit such properties in their exact solutions are primarily 1D
theories (and it is then argued that low dimensionality is essential
to such effects and 2D theories should be more susceptible than 3D
theories).  The discovery of significant $c$-axis effects in the
pnictides  suggests that high-temperature unconventional
superconductivity may not depend in an essential way on low spatial
dimensionality. Since the approach described here is not restricted
to 2D physics, it allows the possibility that both cuprates and
pnictides could be described by the same basic physics, with
differences like $c$-axis effects absorbed into differences in
effective interactions, thus influencing only the parameters of the
theory and not its fundamental structure.

\subsection{Multiband, Momentum-Dependent SU(4)}

As we have seen, by introducing the effective creation and
annihilation operators defined in Eq.~(\ref{oneBandEff}) the SU(4)
model can be reduced to its original form in which
momentum-dependent effects have been averaged over.  This is a
reasonably good approximation for the cuprates where one deals with
effectively single-band physics and a gap that is nodal but a Fermi
surface that is topologically connected.  In the iron arsenides the
Fermi surface is more complex, with disconnected sheets in different
regions of the Brillouin zone. This implies that averaging over $k$
may miss qualitatively important physics for measurements that
resolve momentum.  In Ref.~\cite{sun07} we extended the SU(4) model
to include explicit $k$ dependence for a single band.  We may
generalize this SU(4)$_k$ model to include multiple bands and gaps
in the following way.

Since the pairing formfactor $g(\bm k)$ remains uncertain in the
iron arsenides and in the most general case might even differ from
compound to compound, we take it initially as unknown, to be
determined by measurements.  We then generalize the $k$-dependent
SU(4) formalism of Ref.~\cite{sun07} by introducing
momentum-dependent coupling strengths $G_{kk'}^i$ for pairing and
$\chi_{kk'}$ for antiferromagnetism through
\begin{equation}
 G_{kk'}^i = G_i^0 g_k g_{k'} \qquad
\chi_{kk'} = \chi^0 g_k g_{k'} ,
\label{couplingStrengths}
\end{equation}
where $g_k \equiv |g(\bm k)|$ and $G_i^0$ and $\chi^0$ are
parameters that are independent of momentum but may depend on doping
and temperature. This expression for $\chi_{kk'}$ is identical to
that of Eq.~(14) of Ref.~\cite{sun07} but the expression for
$G_{kk'}^i$ generalizes Eq.~(13) of Ref.~\cite{sun07}, in
anticipation of a richer pairing structure because of the multiband
physics that we expect to be important in the FeAs compounds.

We may now develop the formalism in a manner parallel to that
described in Refs.~\cite{sun07,guid99}.  One obtains a set of gap
equations that generalize the BCS gap equations. Introducing a
doping parameter $x$ and defining a critical doping $x_q$ by
\begin{equation}
 x_q = \sqrt{\frac{\chi-G_0}{\chi-G_1}},
\label{critDope}
\end{equation}
where $G_i = G_i^0 \bar g^2$, $\chi = \chi^0 \bar g^2$, and $\bar g$
is an averaged $g_k$, the solutions of the gap equations for
temperature $T=0$ and momentum $\bm k$ if  $x \le x_q$ are found to
be
\begin{subequations}
\label{gapsUnderdoped}
\begin{eqnarray}
   \singletGap (\bm k) \mwgalign= \displaystyle\frac\Omega2 G_0 \frac{g(\bm k)}{\bar g}
                    \sqrt{x(x_q^{-1} -x)}
\label{gapsUnderdoped1}
\\[8pt]
   \tripletGap (\bm k) \mwgalign= \displaystyle\frac\Omega2 G_1 \frac{g(\bm k)}{\bar g}
                    \sqrt{x(x_q -x)}
\label{gapsUnderdoped2}
\\[8pt]
   \Delta_q(\bm k) \mwgalign= \displaystyle\frac\Omega2 \chi \frac{g(\bm k)}{\bar g}
                    \sqrt{x(x_q^{-1} -x)(x_q-x)}
\label{gapsUnderdoped3}
\\[8pt]
   \lambda_k^\prime  \mwgalign= -\displaystyle\frac\Omega2 \frac{g(\bm k)}{\bar g}
     \left[
        (\chi-G_1)x_q(1-x_qx)+G_1x
     \right]
\label{gapsUnderdoped4}
\end{eqnarray}
\end{subequations}
% \begin{equation}
%  \begin{array}{rcl}
%    \singletGap (\bm k) \mwgalign= \displaystyle\frac\Omega2 G_0 \frac{g(\bm k)}{\bar g}
%                     \sqrt{x(x_q^{-1} -x)}
% \\[8pt]
%    \tripletGap (\bm k) \mwgalign= \displaystyle\frac\Omega2 G_1 \frac{g(\bm k)}{\bar g}
%                     \sqrt{x(x_q -x)}
% \\[8pt]
%    \Delta_q(\bm k) \mwgalign= \displaystyle\frac\Omega2 \chi \frac{g(\bm k)}{\bar g}
%                     \sqrt{x(x_q^{-1} -x)(x_q-x)}
% \\[8pt]
%    \lambda_k^\prime  \mwgalign= -\displaystyle\frac\Omega2 \frac{g(\bm k)}{\bar g}
%      \left[
%         (\chi-G_1)x_q(1-x_qx)+G_1x
%      \right]
%  \end{array}
% \label{gapsUnderdoped}
% \end{equation}
and the corresponding solutions for $x > x_q$ are
\begin{subequations}
\label{gapsOverdoped}
\begin{eqnarray}
   \Delta_q(\bm k) \mwgalign=
   \tripletGap (\bm k) = 0
\label{gapsOverdoped1}
\\
   \singletGap (\bm k) \mwgalign= \displaystyle\frac\Omega2 G_0 \frac{g(\bm k)}{\bar g}
                    \sqrt{1-x^2}
\label{gapsOverdoped2}
\\
   \lambda_k^\prime  \mwgalign= -\displaystyle\frac\Omega2 \frac{g(\bm k)}{\bar g}
     \, G_0 x.
\label{gapsOverdoped3}
\end{eqnarray}
\end{subequations}
Physically, $\singletGap$ and $\tripletGap$ correspond to
correlation energies for singlet and triplet pairing, respectively,
$\Delta_q$ corresponds to correlation energy in the pseudogap state
that is fluctuating AF in nature, and $\lambda'$ denotes the
chemical potential.

These solutions may then be used to determine other physically
important quantities using the methods described in
Refs.~\cite{guid04,guid99,sun05,sun06,sun07,guid07,guid08b,guid08c}.
For example, the superconducting transition temperature $T\tsub c$
is
\begin{equation}
   T\tsub c (\bm k) = G_0 \frac{g(\bm k)}{\bar g} \Omega
        \frac{Rx}{4 k\tsub B \atanh(x)},
\label{Tc}
\end{equation}
where the parameter $R$ is of order one and defined in
Ref.~\cite{sun07}, and because of the AF interaction there are
pseudgap correlations that extend from $T\tsub c$ up to a pseudogap
temperature
\begin{equation}
  T^* (\bm k) = \chi \frac{g(\bm k)}{\bar g} \Omega
        \frac{R(1-x^2)}{4 k\tsub B}.
\label{Tstar}
\end{equation}
The pseudogap states lying between $T\tsub c$ and $T^*$ are
correlated by AF and pairing, but are not expected to have large
static order parameters because of the fluctuations discussed in the
earlier section on spatial inhomogeneity.  Physically these states
may be interpreted in terms of competing AF and SC correlations, but
in a paired basis.  Thus they unify the preformed pair and competing
order pictures for pseudogap states.

Equations (\ref{couplingStrengths})--(\ref{Tstar}) define a
$k$-dependent SU(4) model that can accommodate multiband physics.
They are appropriate for comparison with experimental data that can
resolve $\bm k$.  If one averages these expressions over all momenta
$\bm k$ near the Fermi surface then the averaged factors $\langle
g(\bm k)/\bar g \rangle \rightarrow 1$ and
Eqs.~(\ref{couplingStrengths})--(\ref{Tstar}) reduce to the
equations of the original SU(4) model.  These are appropriate for
comparison with experimental quantities that do not resolve $\bm k$.

\subsection{Multiple Pairing Gaps in the FeAs Compounds}

The gap equations and their solutions given in the preceding section
represent a general formalism applicable for systems in which the
pairing involves multiple bands and the possibility of multiple
pairing gaps within the Brillouin zone. Let us now apply this
formalism specifically to an analysis of the iron superconductors.

We have argued that consistency of pairing with observed
antiferromagnetic structure strongly constrains the permissable
forms of the pairing formfactor $g(\bm k)$ and that the forms most
consistent with the observed properties of the FeAs compounds are
$g(\bm k) = \cos k_x \cos k_y$ or $g(\bm k) = \sin k_x \sin k_y$.
Let us first consider the $\cos k_x \cos k_y$ case. Restricting
attention for purposes of illustration to doping $x$ less than the
critical value $x_q$, we obtain from Eq.~(\ref{gapsUnderdoped}) a
singlet pairing gap
\begin{equation}
  \singletGap (\bm k) = \Delta_0 \cos k_x \cos k_y
   \qquad
  \Delta_0 \equiv \frac{G_0\Omega}{2\bar g}
   \sqrt{x(x_q^{-1} -x)}.
\label{gapcoskxcosky}
\end{equation}
ARPES measurements of Ref.~\cite{nak08} find evidence for a total of
four sheets of Fermi surface within the Brillouin zone.  To
illustrate ideas in a transparent way, let us introduce a simple
model in which we assume the four pockets of Fermi surface (labeled
$\alpha$, $\beta$, $\gamma$, and $\delta$) to be spheres centered at
the appropriate momentum, with the radii  $k_\alpha$, $k_\beta$,
$k_\gamma$, and $k_\delta$ of the
spheres determined by fits to the ARPES data.  We illustrate in Fig.~\ref{fig:FeAsSphericalFermiSurfaces}.%
\singlefig
{FeAsSphericalFermiSurfaces}       % Label
{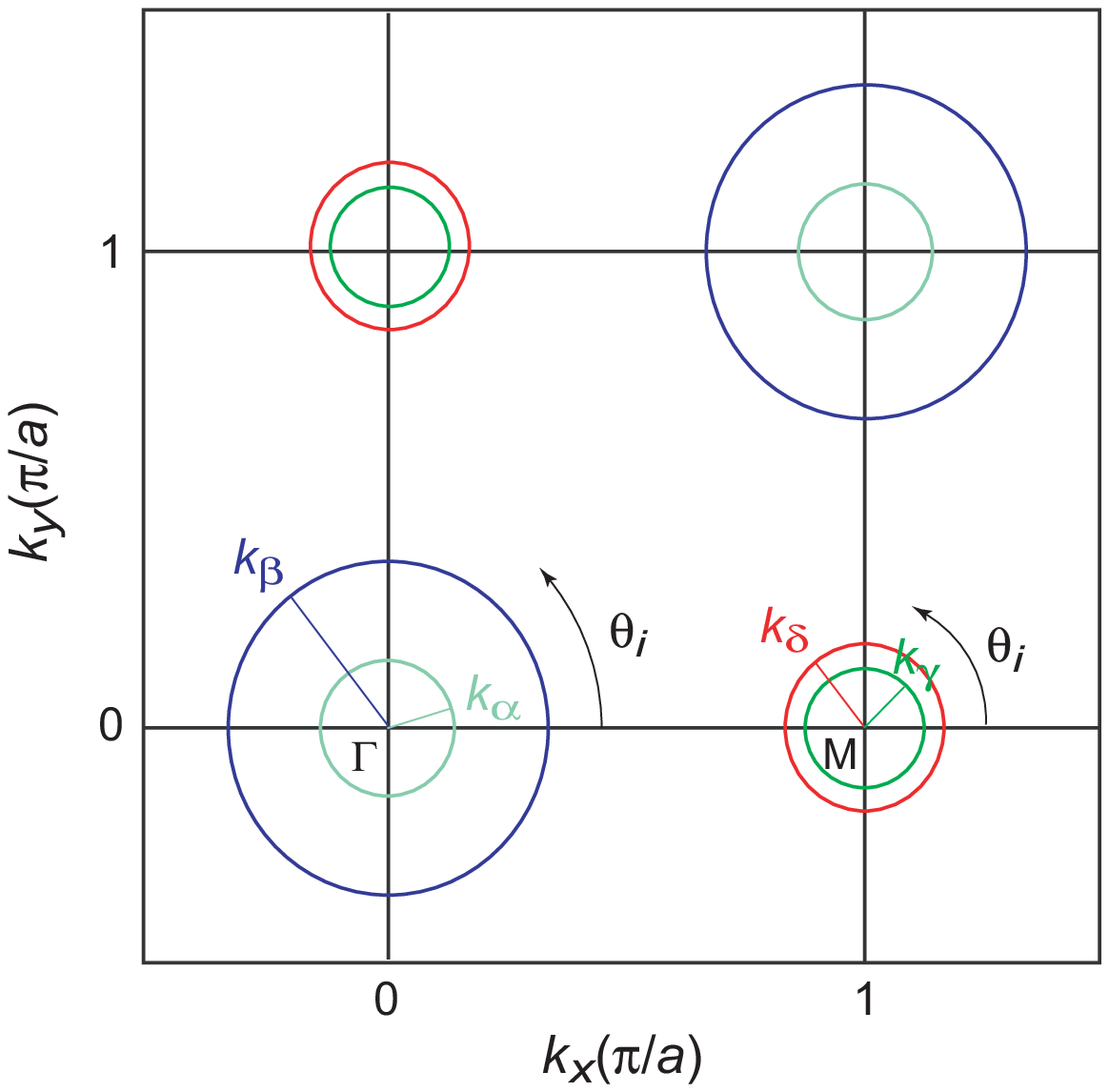}    % Filename (with path if needed)
{0pt}         % Extra space above (with units)
{0pt}         % Extra space below (with units)
{0.5}         % Scale factor (1.0 is full size)
{Spherical approximations to the Fermi surfaces. The four Fermi
surface pockets, two around the $\Gamma$ point and two around the X
point are approximated by circles.}
Then from Eq.~(\ref{gapcoskxcosky}) the pairing gaps on the four
sheets of Fermi surface are given by
\begin{equation}
  \Delta_i \equiv \singletGap (k_i, \theta_i) = \Delta_0
     \cos(k_i \cos \theta_i) \cos(k_i \sin \theta_i),
\label{gaps4sheets}
\end{equation}
where $i= \alpha, \beta, \gamma, \delta$ labels the Fermi surface
sheets and the polar angles $\theta_i$ are centered at the $\Gamma$
and $M$ points of the Brillouin zone (see
Fig.~\ref{fig:FeAsSphericalFermiSurfaces}). Writing this out
explicitly for the four cases we obtain the results given in
Table \ref{FeGapTable2} for the gaps on the four pockets of Fermi surface.%
%
%
% ---------------  Beginning of Table  --------------- %
%
{\renewcommand\arraystretch{1.20} % vertical stretch factor
\begin{table}[t]
  \centering
  \caption{Pairing gaps on four sheets of the idealized Fermi surface.}
  \label{FeGapTable2}
\vspace{1pt}
  \begin{normalsize}
    \begin{centering}
      \setlength{\tabcolsep}{5 pt}
      \begin{tabular}{ccc}
        \hline
            Label &
            Fermi surface &
            Pairing gap

        \\        \hline
            $\Delta_\alpha$ &
            $k_x^2 + k_y^2 = k_\alpha^2$ &
            $\Delta_0 \cos(k_\alpha \cos \theta_\alpha)
           \cos(k_\alpha \sin \theta_\alpha)$

        \\[1 pt]
            $\Delta_\beta$ &
             $k_x^2 + k_y^2 = k_\beta^2$&
            $ \Delta_0 \cos(k_\beta \cos \theta_\beta)
           \cos(k_\beta \sin \theta_\beta)$

        \\[1 pt]
            $\Delta_\gamma$ &
             $k_x^2 + k_y^2 = k_\gamma^2$&
             $\Delta_0 \cos(k_\gamma \cos \theta_\gamma)
           \cos(k_\gamma \sin \theta_\gamma)$

        \\[1 pt]
            $\Delta_\delta$ &
            $k_x^2 + k_y^2 = k_\delta^2$ &
            $\Delta_0 \cos(k_\delta \cos \theta_\delta)
           \cos(k_\delta \sin \theta_\delta)$
        \\[1 pt]        \hline
      \end{tabular}
    \end{centering}
  \end{normalsize}
\end{table}
}
%
% ---------------  End of Table  --------------- %
%

A fit of the parameters $\Delta_0$ and the $k_i$ to the data of
Ref.~\cite{nak08} gives the description of the pairing gaps
illustrated in
Figs.~\ref{fig:4gaps}--\ref{fig:gapVscoskxcosky},%
\singlefig
{4gaps}       % Label
{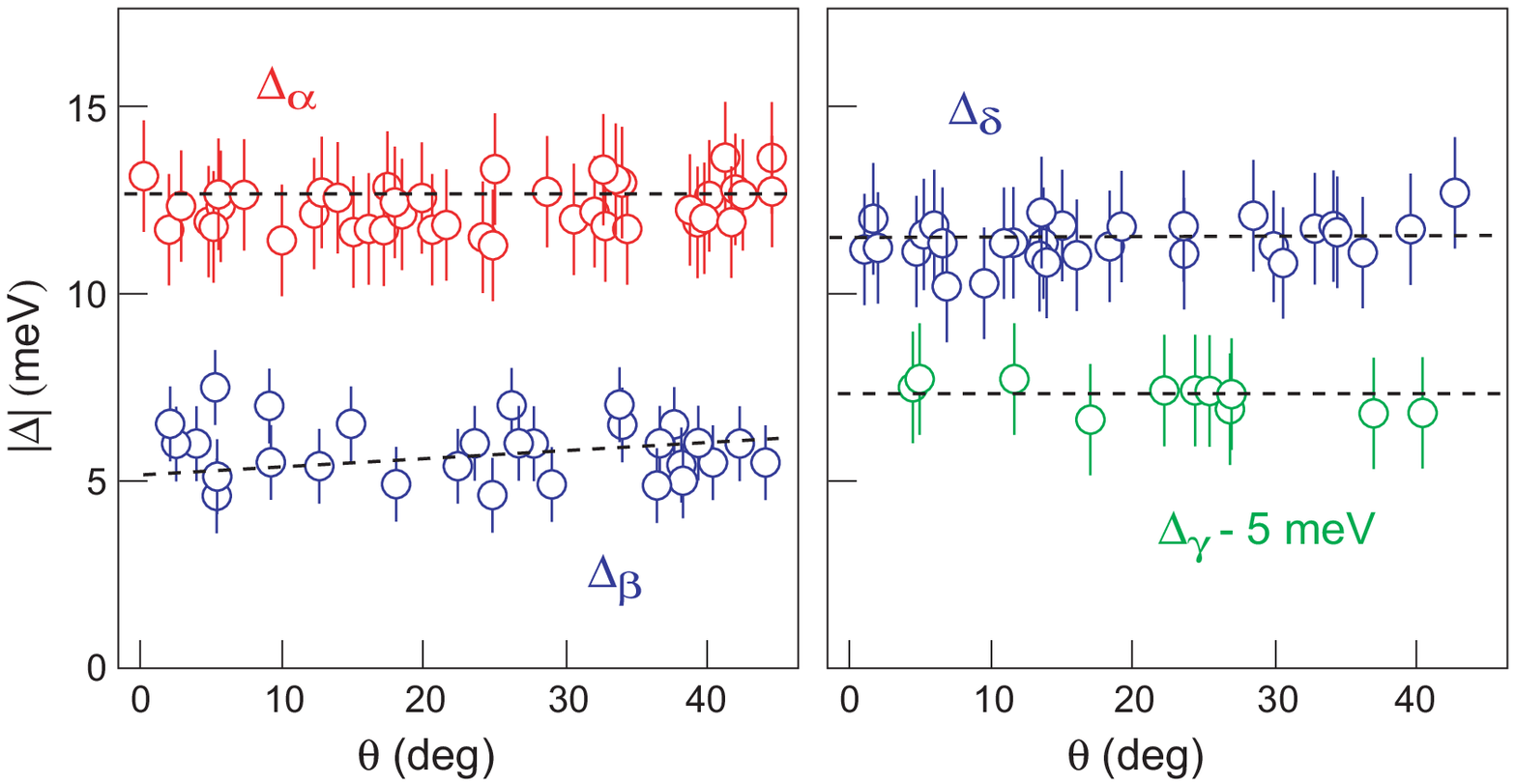}    % Filename (with path if needed)
{0pt}         % Extra space above (with units)
{0pt}         % Extra space below (with units)
{0.49}         % Scale factor (1.0 is full size)
{Pairing gaps on four sheets of the Fermi surface. Note that
$\Delta_\gamma$ is displaced by 5 meV for plotting purposes. Circles
are data from Ref.~\cite{nak08} and dashed lines are theoretical
using Eq.~(\ref{gapcoskxcosky}). Parameters $\Delta_0 = 13.5$ meV,
$k_\alpha/\pi = 0.135$, $k_\beta/\pi = 0.370$, $k_\gamma/\pi =
0.141$, and $k_\delta/\pi = 0.181$ were determined by fitting to the
data.}
\singlefig
{FSoverlay}       % Label
{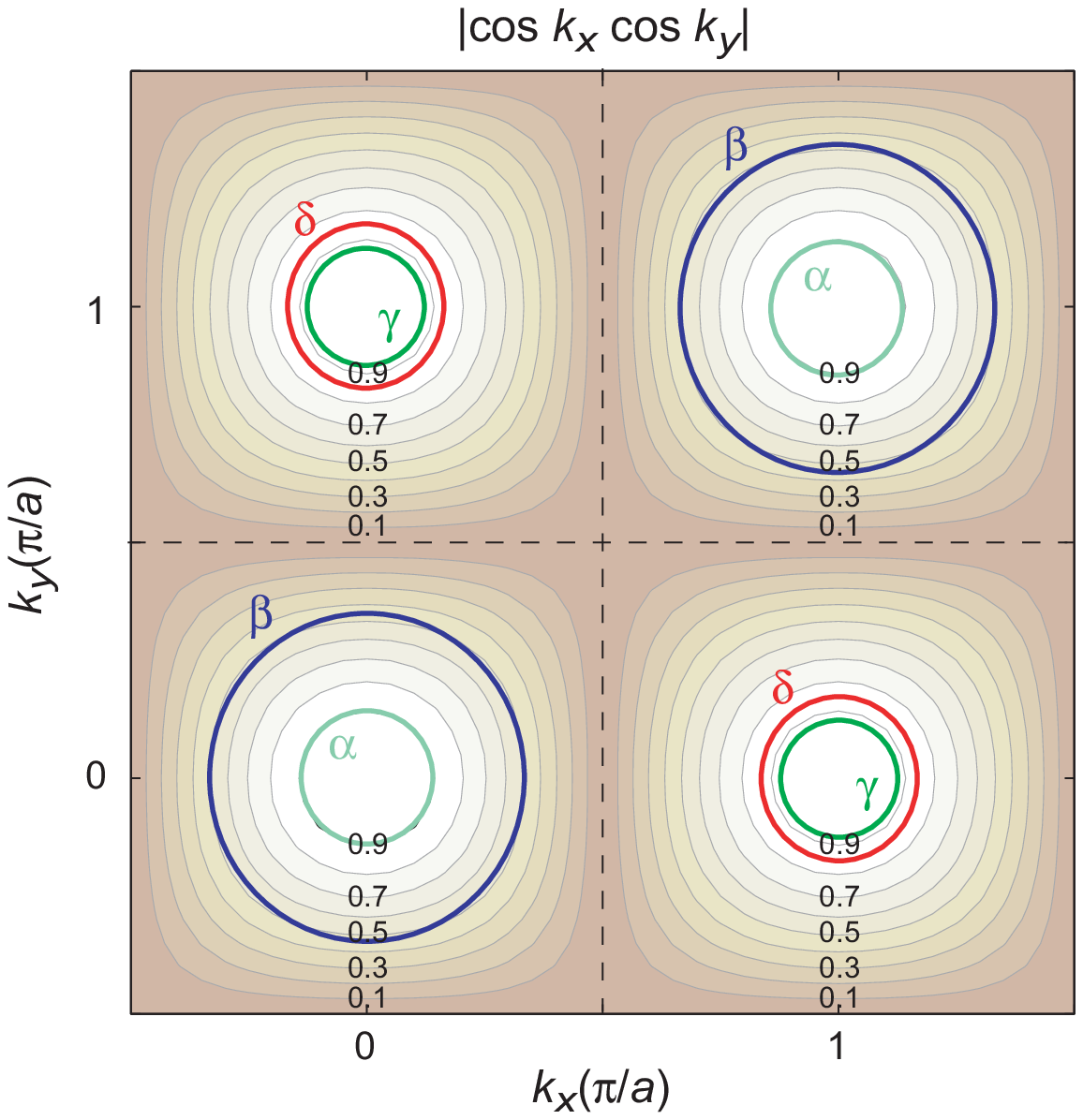}    % Filename (with path if needed)
{0pt}         % Extra space above (with units)
{0pt}         % Extra space below (with units)
{0.53}         % Scale factor (1.0 is full size)
{Fermi surfaces $\alpha$, $\beta$, $\gamma$, and $\delta$ superposed
on contours of the pairing formfactor $|\cos k_x \cos k_y|$.  Gap
nodes are indicated by dashed lines.}
\singlefig
{gapVscoskxcosky}       % Label
{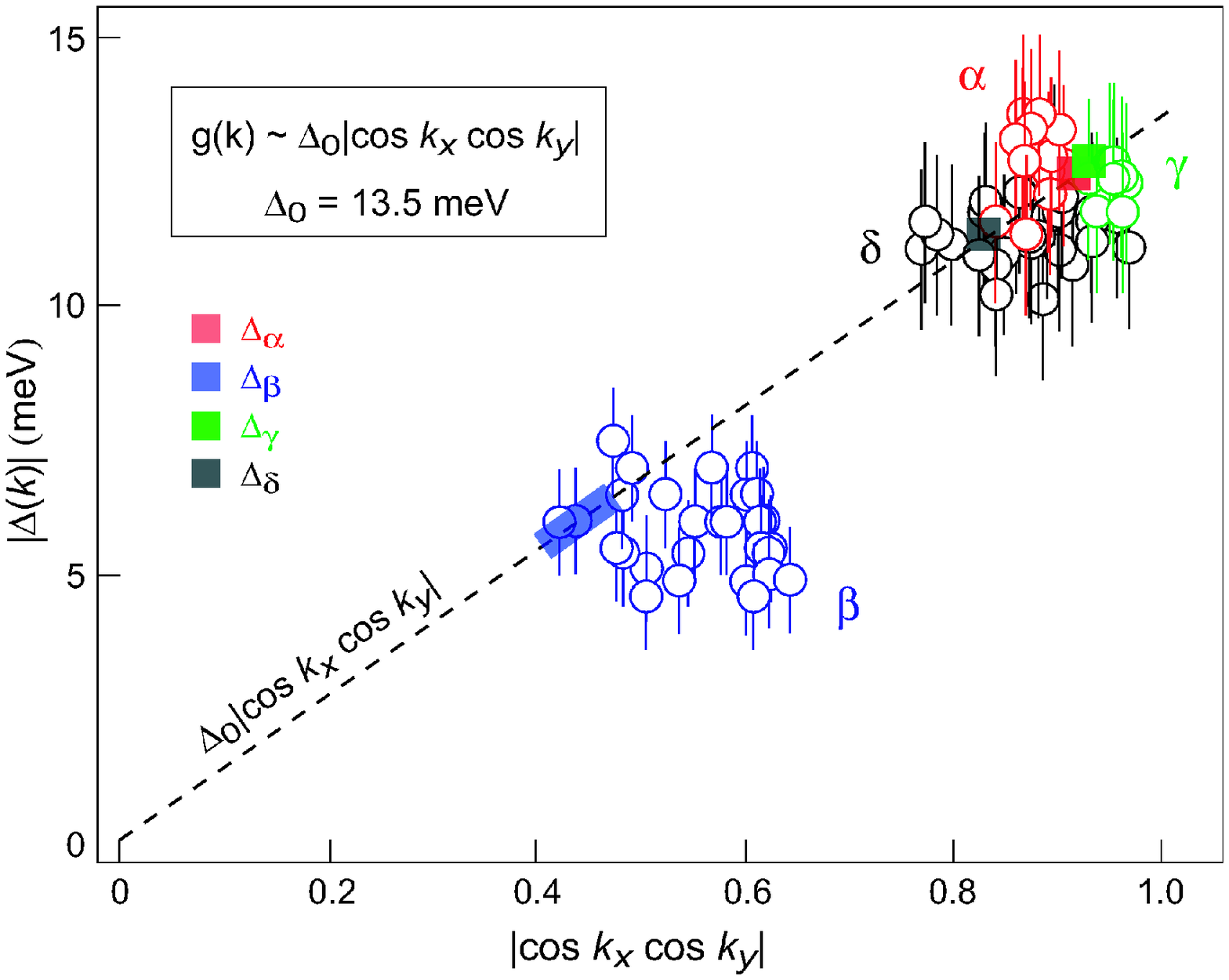}    % Filename (with path if needed)
{0pt}         % Extra space above (with units)
{0pt}         % Extra space below (with units)
{0.45}         % Scale factor (1.0 is full size)
{Pairing gaps on four sheets of the Fermi surface versus $|\cos k_x
\cos k_y|$.  Data from Ref.~\cite{nak08} and the squares and
rectangles indicate theoretical values for gaps on the four sheets
deduced from Fig.~\ref{fig:FSoverlay}. The same parameters as for
Fig.~\ref{fig:4gaps} were used.}
where in Fig.~\ref{fig:4gaps} the circles indicate data with
associated uncertainties given by the error bars and the dashed
lines  represent the gaps calculated from Eq.~(\ref{gaps4sheets}).
Thus we see that the ARPES measurements of Ref.~\cite{nak08} are at
least approximately consistent with the $\cos k_x \cos k_y$ pairing
gap formfactor deduced in the present paper by requiring
self-consistency of antiferromagnetism and superconductivity within
an SU(4) symmetry.

The $\sin k_x \sin k_y$ formfactor that would also be compatible
with the antiferromagnetism according to Table \ref{FeGapTable}
would on the other hand not be compatible with the data displayed in
Fig.~\ref{fig:4gaps}, since it would imply nodes on the Fermi
surfaces (compare Figs.~\ref{fig:contour4filled} and
\ref{fig:FSoverlay}) not observed in the data.  We conclude that
consistency of the observed antiferromagnetism with the
superconductivity in the FeAs compounds is possible with either
$\cos k_x \cos k_y$ or $\sin k_x \sin k_y$ pairing formfactors, but
requiring in addition consistency with the ARPES data of
Ref.~\cite{nak08} restricts to the  $\cos k_x \cos k_y$ choice.

\subsection{Top-Down Approach to Superconductivity}

The approach advocated in this paper may be termed ``top-down'':
The appropriate theory for describing the superconductivity is
inferred from the global properties exhibited by the physical
interacting system, rather than  from the underlying atomic and
crystal structure of the non-interacting system (which we shall term
``bottom-up'' approaches). This is not the most common approach to
this problem, but it is a powerful one. It is closely related to the
philosophy advocated in emergent theories of many-body structure
where it is argued that the appropriate building block for a
theoretical understanding of complex systems are the ones actually
observed to occur, and that these should be considered just as
fundamental as more ``microscopic'' building blocks.

Practically, the top-down approach may be on firmer ground than
bottom-up approaches because it is constrained more directly by
data. However, it is not phenomenological since it is based on an
exact many-body solution within a truncated space.  Therefore,
comparing a top-down analysis with data constrains the validity of
proposed bottom-up models. A specific example is afforded by the
analysis associated with Table \ref{FeGapTable}, where neutron
scattering data were shown to restrict severely the collective
pairing structures that can be consistent with the observed
competing antiferromagnetism.

Our dynamical symmetry approach thus unifies  cuprate and FeAs superconductivity  in a coherent picture that demonstrates how the Cooper instability originates from a single theoretical framework in these diverse systems.  That the interaction responsible for exploiting the Cooper instability is different in BCS superconductors and cuprates superconductors, and could be different yet again in FeAs superconductors
% (and even more broadly, is certainly different in the variety of superconducting behavior observed in complex atomic nuclei),
is of secondary importance. Seen at the level of abstraction implied
by the non-abelian superconductor hypothesis, these are all the {\em
same superconductors,} all made possible by a Cooper instability
that morphs into more complex behavior but persists in clearly
identifiable form even in the presence of competing degrees of
freedom like antiferromagnetism, and that leads to a phenomenology
with many common features, even when enabled by physically very
different weakly attractive interactions \cite{guid08b}.

Finally, if our identification of FeAs superconductivity as the
second example of non-abelian superconductivity is valid, there may
be experimental lessons to be learned.  The dynamical symmetry
framework presented here unifies cuprate and iron arsenide
superconductivity.  It is then instructive to recall that both the
discovery of cuprate superconductivity and that of FeAs
superconductivity were surprising when viewed from standard
perspectives.  The reason for that surprise lies largely in what we
now realize is a too-narrow view of the microscopic conditions under
which superconductivity could arise. This suggests that one should
search for new examples of non-abelian superconductivity guided not
only by microscopic considerations but also by general principles of
where we might expect the Cooper instability in the presence of
emergent collective degrees of freedom.

The cuprate and iron arsenide data suggest that any region having
magnetism competing with a pairing interaction (which might be
mediated by a variety of possible microscopic interactions) is
fertile ground for such a search, but this is not necessarily the
only possibility. Any strongly collective mode that can compete with
pairing for available strength within a relevant Hilbert space could
lead to non-abelian superconductivity and the attendant complex
behavior exemplified by cuprate and FeAs superconductors, with the
non-abelian algebra not necessarily SU(4) in that case but with the
basic ideas qualitatively similar to those discussed here.

\subsection{Conclusions}

We have presented evidence that the new Fe-based high-temperature
superconductors represent the second example (after the cuprates) of
the non-abelian superconductors proposed in an earlier paper.  These
superconductors differ from normal ones because the non-abelian
properties exhibited by commuting their minimal sets of physical
operators imply  non-linear constraints for collective degrees of
freedom interacting with the superconductivity. The identification
of non-abelian superconductivity in these two classes of compounds
permits a unified model of cuprate and Fe-based superconductors to
be constructed based on an SU(4) group (and subgroups) generated by
emergent degrees of freedom.

The requirement that the SU(4) algebra both close under commutation
and be consistent with the magnetic structure inferred from neutron
scattering experiments permits constraints to be placed on orbital
symmetries for the pairing gap in FeAs compounds.  We find that
neither $s_{x^2+y^2}$ nor $d_{x^2-y^2}$ symmetries appear compatible
with the neutron scattering data but $s_{x^2y^2}$ or $d_{xy}$ could
be, and comparing the predicted gaps with ARPES data restricts the
choice uniquely to $d_{x^2y^2}$ (that is, $\cos k_x \cos k_y$). Thus
we reach the quite interesting conclusion that a unified SU(4) model
of FeAs and cuprate high temperature superconductivity is possible,
but consistency with neutron scattering and ARPES data requires that
the pairing in the two cases corresponds to {\em different} orbital
formfactors at the microscopic level.

\acknowledgments We thank Pengcheng Dai for useful discussions
concerning neutron scattering on the Fe-based superconductors and
Adriana Moreo for illuminating comments on the general physics of
FeAs superconductors.

\vfill

\end{document}